\documentclass[12pt,preprint]{aastex}

\slugcomment{accepted by AJ: 16 December, 2005}

\shorttitle{Disk around HD169142}
\shortauthors{Raman, Lisanti et al. }

\begin{document}

\title{A Keplerian Disk around the Herbig Ae star HD169142}

\author{A. Raman\altaffilmark{1}, 
        M. Lisanti\altaffilmark{1,2}, 
        D.J. Wilner\altaffilmark{1}, 
        C. Qi\altaffilmark{1},
        M. Hogerheijde\altaffilmark{3}
        }

\altaffiltext{1}{Harvard-Smithsonian Center for Astrophysics,
  60 Garden Street, Cambridge, MA 02138, USA}

\altaffiltext{2}{Stanford University Department of Physics, 
382 Via Pueblo Mall, Stanford, CA 94305-4060}

\altaffiltext{3}{Leiden Observatory, P.O. Box 9513, 2300 RA, Leiden, 
  The Netherlands}

\begin{abstract}
We present Submillimeter Array observations of the Herbig Ae star 
HD169142 in 1.3~millimeter continuum emission 
and $^{12}$CO J=2-1 line emission at $\sim1\farcs5$ resolution
that reveal a circumstellar disk.
The continuum emission is centered on the star position and resolved, 
and provides a mass estimate of $\sim0.02$~$M_{\odot}$ for the disk. 
The CO images show patterns in position and velocity that are well matched
by a disk in Keplerian rotation with low inclination to the line-of-sight. 
We use radiative transfer calculations based on a flared, passive 
disk model to constrain the disk parameters by comparison to
the spectral line emission. The derived disk radius is $235$~AU, 
and the inclination is $13^{\circ}$.
The model also necessitates modest depletion of the CO molecules, 
similar to that found in Keplerian disks around T Tauri stars.
\end{abstract}

\keywords{stars: individual(HD169142) --- 
          stars: pre-main sequence ---
          stars: planetary systems: protoplanetary disks
}

\section{Introduction}
Millimeter-wave observations of disks surrounding pre-main sequence stars
are a promising probe of the planet formation process \citep{bec96}.
While most studies have concentrated on the circumstellar disks around
low mass T Tauri stars, far less is known about the 
disks around their intermediate mass counterparts, the Herbig Ae/Be stars.
Better characterization of the circumstellar disks around these higher mass 
stars will help address how stellar mass affects the development of 
planetary systems.

Detailed modeling of the spectral energy distributions of Herbig Ae/Be
stars has generally supported the protoplanetary disk paradigm
\citep[e.g.][]{hil92,chi01,dom03}.
But out of the large sample of Herbig Ae/Be star disk candidates
recognized by infrared excess from circumstellar dust \citep{the94,mal98},
only a handful have been spatially and spectrally resolved through 
direct imaging observations to show clear signatures of rotationally
supported disks of Solar System size \citep{man97,man00,pie03,cor05}.

The emission line star HD169142 (SAO 186777, MWC 925; spectral type A5)
with infrared excess (IRAS 18213-2948) was included in the \citet{the94}
catalog of 287 Herbig Ae/Be candidates. Subsequent photometric
and spectroscopic observations filled out its spectral energy distribution
at infrared and submillimeter wavelengths \citep{syl96,dun97,mal98}.
At $\sim145$~pc distance, HD169142 is notably isolated from any molecular cloud
or extensive reflection nebulosity, though \cite{kuh01} used differential
polarimetry to detect a $\sim200$~AU ($\sim1\farcs5$) diameter 
circular halo of scattered light. While some doubt lingers as to the 
pre-main-sequence nature of HD169142, it is among isolated Herbig Ae 
stars selected by \cite{mee01} and classified as ``group Ib'' with a 
strong mid-infrared component, and a conspicuous absence of 10~$\mu$m 
silicate emission \citep{bou01}. This star also has prominent 
polycyclic aromatic hydrocarbon emission features \citep{hab05}.

The spectral energy distribution of HD169142 is well fit by
a passive irradiated disk model, though the near-infrared flux seems low, 
perhaps because the inner disk rim is viewed pole-on \citep{dom03}, 
or perhaps because extensive inner disk clearing is underway,
possibly together with planet formation.  The relative isolation of 
this star from any molecular clouds may indicate an advanced 
pre-main-sequence age. 
A circumstellar molecular component is detected via relatively strong 
emission in the CO J=3-2 line, with the narrow CO line profile modeled 
with a near face-on disk \citep{gre00,den05a}. 

Here we present high resolution observations of HD169142 from the 
Submillimeter Array\footnote{The Submillimeter Array 
is a joint project between the Smithsonian Astrophysical Observatory and 
the Academia Sinica Institute of Astronomy and Astrophysics, and is
funded by the Smithsonian Institution and the Academia Sinica.}
(SMA) that clearly resolve a circumstellar disk in Keplerian rotation.
We derive the basic disk properties from arcsecond resolution images of 
the $^{12}$CO J=2-1 line using radiative transfer calculations based 
on a flared, passive disk model \citep{dal05}.

\section{Observations}
We used the SMA \citep{ho04} to observe HD169142 on 19 April 2005.
Table~\ref{tab:obs} lists the observational parameters.
Seven antennas were available in an extended configuration 
that provided a baseline range of 27 to 181 meters and a 
synthesized beam of $1\farcs6 \times 1\farcs0"$ at position angle $20^{\circ}$. 
The target source was tracked over the hour angle range -1.0 to 3.5. 
Weather conditions were excellent, and system temperatures 
ranged from 85 to 150 K (DSB).
The correlator provided 2 GHz of bandwidth in each sideband 
and was configured to include the 
$^{12}$CO J=2-1 line at 230.538 GHz in the upper sideband 
in a 104 MHz spectral ``chunk'' with channel spacing 0.2 MHz 
($\sim$0.27 km s$^{-1}$).
Complex gain calibration was performed using the 
calibrator sources J1911-201 and J1924-292. 
Passband calibration was done with observations of 
the strong sources 3c279 and Uranus.  
The absolute flux scale was set using observations of Callisto 
and is estimated to be accurate to about 10\%.
All of the calibration was done using MIR software,
followed by standard imaging and visual analysis using the miriad package.

\section{Results and Discussion}

\subsection{Continuum}
\label{sec:continuum}

The lower right panel of Figure~\ref{fig:co21+cont}a shows the 
continuum image made from the upper sideband visibility data. 
Fitting an elliptical Gaussian to the visibilities gives 
a flux of $169\pm5$ mJy, and a fwhm size of $\sim0\farcs9$.
The fit does not indicate any significant ellipticity. 
The observed flux is consistent with the previous 
single dish 1.3~mm measurement of $197\pm15$~mJy \citep{syl96}
taking account of the uncertainties, which suggests that 
the interferometer is not missing any large scale emission.
The 1.3~mm continuum emission arises from a compact region, 
which we associate with the circumstellar disk. The spectral
energy distribution indicates that thermal dust emission strongly
dominates at this wavelength; any contribution from ionized gas 
emission is negligible.

The dust continuum flux, $S_{\nu}$, may be used to estimate the 
disk mass, $M_{disk}$, assuming optically thin emission, i.e.
$M_{disk} =   { S_{\nu} D^2 }/
             { \kappa_{\nu} B_{\nu}(\langle T \rangle) } $
where $B_{\nu}$ is the Planck function.
We adopt the mass opacity of \cite{bec90}, 
$\kappa_{\nu}$ = 0.02(1.3 mm/$\lambda$) cm$^{2}$~g~$^{-1}$,
which includes a gas-to-dust ratio of 100.
Assuming an average temperature $\langle T \rangle =30$~K,
appropriate for the bulk of the disk material,
the disk mass is $0.02~M_{\odot}$.
The factor of few uncertainty in $\kappa_{\nu}$ likely
dominates the uncertainty in this mass estimate. 
This mass estimate falls in the middle of the range 
of estimates for protoplanetary disks around T Tauri stars.

\subsection{$^{12}$CO J=2-1}
\label{sec:CO}

\begin{figure}
\epsscale{0.8}
\plotone{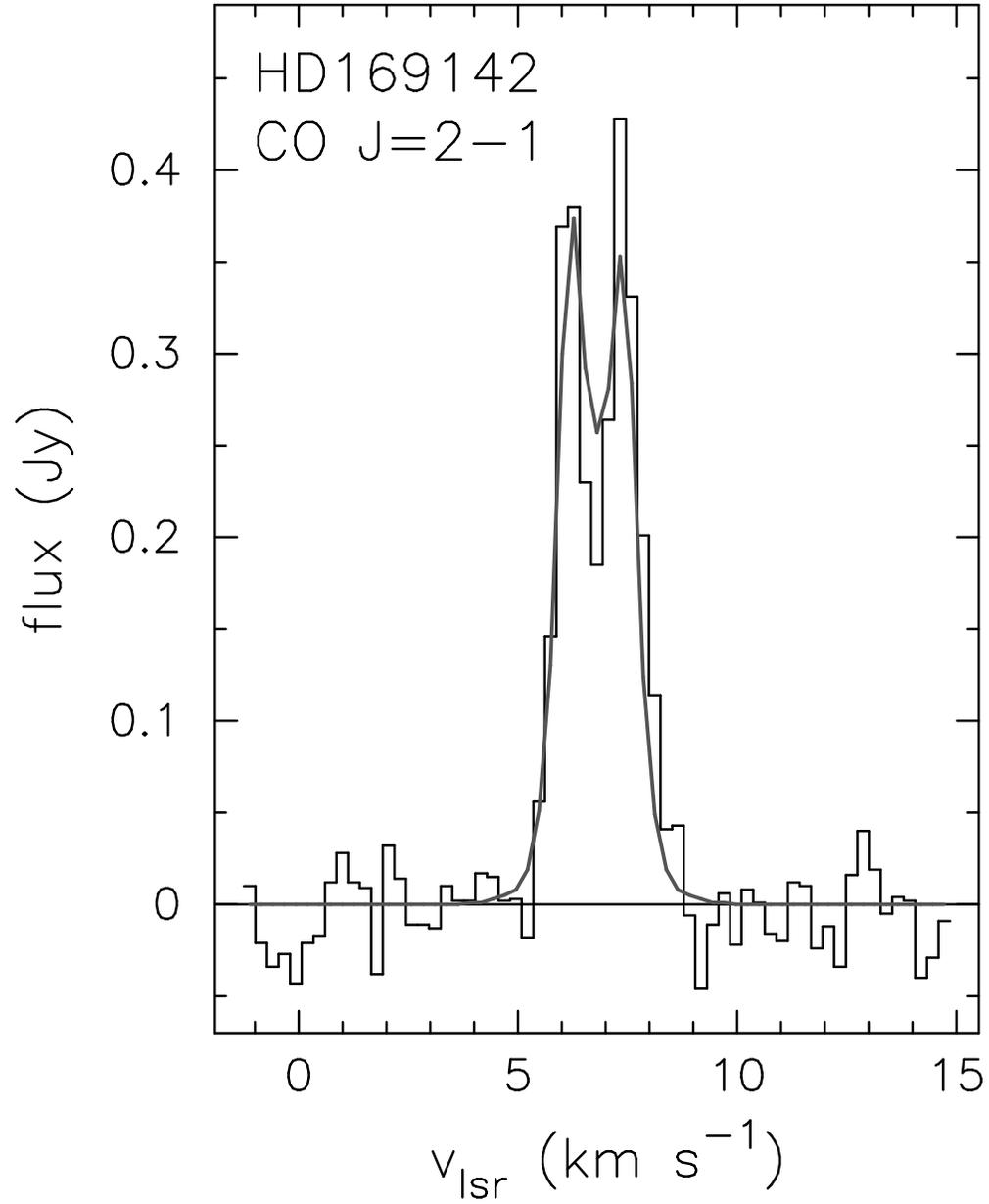}
\caption{
Spectrum of $^{12}$CO J=2-1 line observed from HD169142
at the position of peak continuum emission convolved
to a 3\farcs5 beam size (histogram) 
and the best fit disk model (solid line).
}
\label{fig:co21-spec} 
\end{figure}

Figure~\ref{fig:co21+cont}a shows the CO J=2-1 line emission distribution 
over the velocity range with significant emission.  
The images show an approximately symmetric pattern around the central
velocity of $6.8$~km~s$^{-1}$.  At the extremes, the emission is 
centered at the continuum peak position, while at intermediate velocities
the emission shows a gradient oriented slightly east of north that at 
the central velocities separates into two peaks. 
Figure~\ref{fig:co21-spec} shows the spectrum of the CO J=2-1 line 
convolved to a beam size of $3\farcs5$, which encompasses the full
spatial extent of the emission. The spectral profile is double peaked.
All of these features taken together strongly suggest a disk at 
low inclination to the 
line-of-sight in Keplerian rotation, as calculated by \citet{bec93},
where low J CO lines have high optical depth and trace gas temperature 
near the disk surface.
A Keplerian velocity field is appropriate for the estimated 
low disk mass relative to the central star mass, i.e. 
$M_{disk} (\approx 0.02~M_{\odot}) << M_{star}~(\approx 2.0~M_{\odot})$.

\subsubsection{Disk Model}
In order to verify the Keplerian nature of the observed rotation of
the outer disk material, and to provide quantitative constraints on the 
basic disk parameters, we 
performed a series of radiative transfer calculations based on a 
physical model selected from the catalog of irradiated accretion disk 
models of \citet{dal05}. 
We selected the model that \cite{den05b} found to closely match the
spectral energy distribution of HD169142. In this model, the 
central star parameters are $T_{eff} = 9000$~K, $M_{*} = 2.0~M_{\odot}$, 
$R_{*} = 1.7~R_{\odot}$, and age 10~Myr, which are well matched to 
the stellar properties of HD169142.
The disk has a mass accretion rate of $1.0 \times 10^{-8}~M_{\odot}$~yr$^{-1}$, 
with viscosity parameter, $\alpha = 0.01$. The dust in the disk is well mixed 
with the gas and has a power law size distribution 
$n(a) \sim a^{-p}$, $p = 3.5$, with 
sizes ranging from $a_{min} = 0.005$~$\mu$m and $a_{max} = 1$~mm.

We identified an optimal disk radius, inclination, and orientation on
the plane of the sky, and CO abundance by performing a chi-squared 
minimization, comparing the SMA visibility data and visibilities 
derived from a grid of disk models spanning a range for each parameter.
The radiative transfer calculations were done with 
the 2D accelerated Monte Carlo code of \cite{hog00} to generate simulated 
data cubes from which visibilities were derived at the $(u,v)$ points 
observed by the SMA for the calculation of the chi-squared metric.
The (dust) temperature and density values from the model were re-binned 
and truncated for a range of outer disk radii from 150 to 400 AU, 
based on the extent seen in the CO J=2-1 images. 
Because of beam dilution, these data do not constrain the inner disk radius. 
The narrow line width suggests a 
viewing angle close to face-on for the disk, and we explored a range 
of inclinations from $0^{\circ}$ to $30^{\circ}$.
The turbulent component of the velocity field was fixed at the 
low value of 0.12~km~s$^{-1}$, less than the resolution of the data.
The CO abundance is characterized by a depletion factor from the nominal 
dark cloud abundance of $10^{-4}$ with respect to molecular hydrogen.

\begin{figure}
\epsscale{0.5}
\plotone{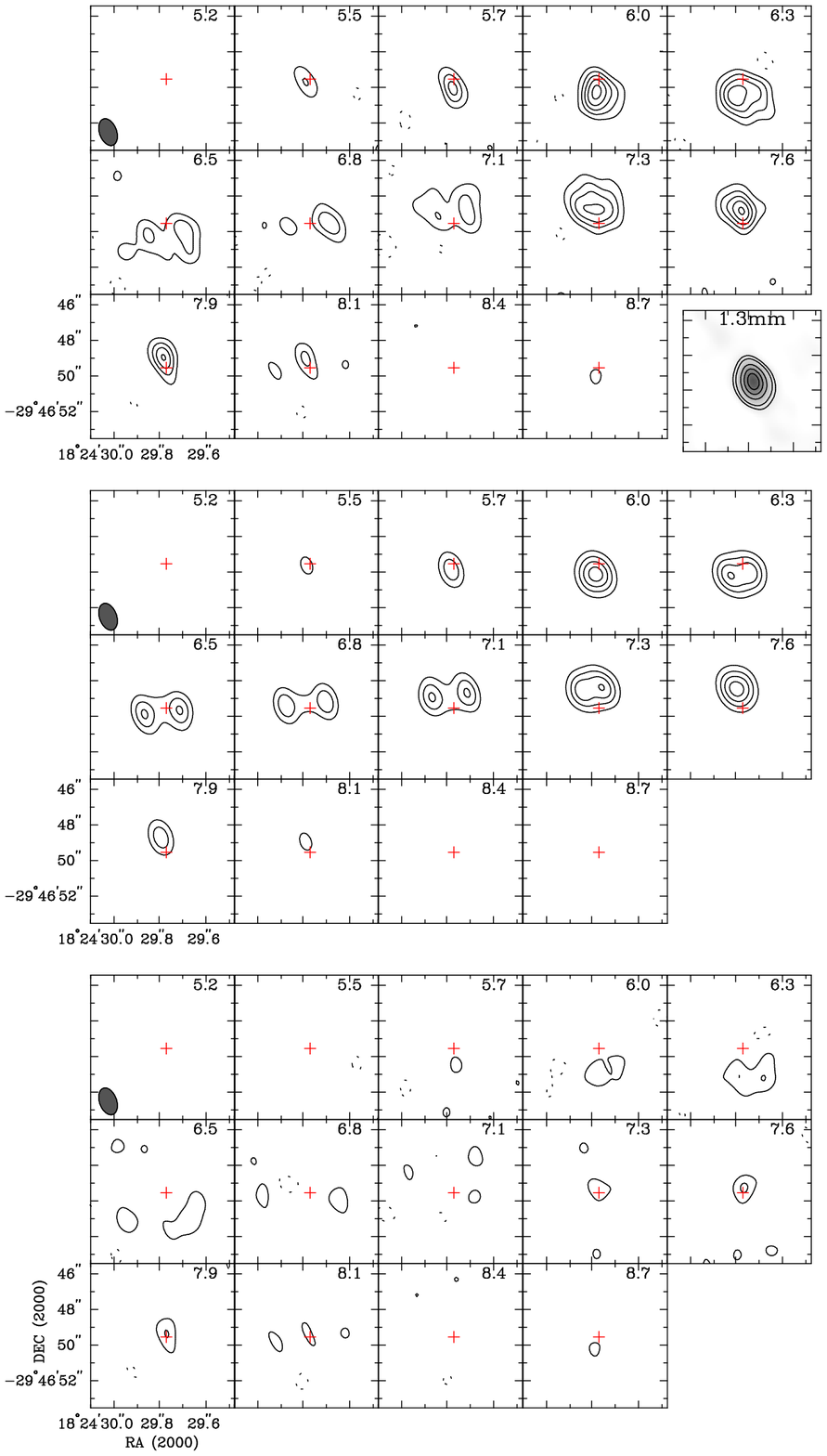}
\caption{
{\em (a) upper:}
A series of $^{12}$CO J=2-1 line velocity images of HD169142,
and the 1.3 millimeter continuum image inset at the bottom right.
The contour levels for the line images are $-2,2,4,6,...\times200$~mJy.
Negative contours are dotted.  The small cross marks the position of the 
continuum peak.
The contour levels for the 1.3 millimeter continuum image 
are $5,10,15,...\times5$~mJy.
The filled ellipse in the lower left corner of the upper left panel shows 
the $1\farcs6\times1\farcs0$ p.a. $20^{\circ}$ synthesized beam.
{\em (b) middle:}
The $^{12}$CO J=2-1 line velocity images for the best fit disk model
(see text). The contour levels are the same as in (a).
{\em (c) lower:}
Difference images formed from the subtraction of the best fit model 
visibilities from the observations. 
The contour levels are the same as in (a).
}
\label{fig:co21+cont} 
\end{figure}

We find a best fit disk model with radius $235\pm 5$~AU and 
inclination $13\pm 1^{\circ}$, oriented at position angle $5\pm 5^{\circ}$,
with a CO abundance of $2.5\times 10^{-5}$ (depletion factor 4).
We assign uncertainties to the fitted parameters that are somewhat 
larger than the formal uncertainties based on visual examination 
of the model images and their residuals.
Table~\ref{tab:model} summarizes the model parameters.
Figure~\ref{fig:co21+cont}b shows the velocity channel maps from 
simulated observations of the best fit model. 
The Keplerian disk model clearly captures the 
main features visible in the data. 
Figure~\ref{fig:co21-spec} shows the spectrum at $3\farcs5$ resolution
derived from the model, overlayed on the data. This comparison 
gives an impression of differences integrated over the full 
spatial extent of disk emission.
The derived low inclination is compatible with the lack of detectable
ellipticity in the 1.3 millimeter dust emission, as well as the 
approximately circular appearance of the polarized scattered 
optical light nebula.
The derived disk outer radius of 235 AU is larger than the 
130~AU inferred from unresolved single dish CO J=3-2 data \citep{den05a}, 
though somewhat smaller than those found for other resolved pre-main-sequence 
A star systems, for example 545~AU for MWC480  \citep{sim00} or
835 AU for HD34282 \citep{pie03}, and comparable
to some lower mass T Tauri star disks \citep{sim00}.
The modestly depleted CO abundance in this disk is similar to values 
typically found for the Keplerian disks around T Tauri stars, for example 
$1.4\times 10^{-5}$ derived for the DM Tau disk by Dutrey et al. (1997) 

The channel maps present minor asymmetries and irregularities that 
are not reproduced by the simple, smooth, Keplerian model. However,
there are no dramatic departures from this equilibrium configuration 
like those seen in the circumstellar material around AB Aur, which is 
perhaps a considerably younger Herbig Ae star \citep{pie05,cor05}.
Figure~\ref{fig:co21+cont}c shows images made from the difference between 
the disk model and the data for each channel. These difference images 
show faint, extended emission in the data at blueshifted velocities, 
and slight excess in central peak brightness at redshifted velocities. 
These differences between the data and the model are not especially 
significant, but they may reflect real source structure.  
Observations with higher sensitivity and higher angular resolution 
will be needed to be more definitive.

\section{Conclusions} 
We have imaged the Herbig Ae star HD169142 in 1.3~millimeter 
continuum emission and $^{12}$CO J=2-1 line emission at 
$\sim1\farcs5$ resolution. 
These observations resolve a disk of molecular gas in Keplerian 
rotation around the star. The presence of the disk is strong evidence 
that HD169142 is a {\it bona fide} pre-main-sequence star, despite its 
distance from any molecular cloud. The star's spectral energy distribution,
isolation, and the regularity of its disk velocity field suggest 
an advanced pre-main-sequence evolutionary state.
The millimeter dust emission indicates a mass of $\sim0.02$~$M_{\odot}$ 
for the disk. Modeling the CO line emission shows that the disk has an 
outer radius of 235~AU and is viewed close to face-on, compatible with 
an associated scattered light nebula seen with differential polarimetry.
The disk mass, size, and CO depletion are in line with those 
found for resolved Keplerian disks around other young stars.

\acknowledgments
We thank Jim Moran, the SMA director, for allocating observing time to 
the Harvard University Astronomy 191 undergraduate class that allowed
A.R. and M.L. to undertake this work. We thank Nimesh Patel for help 
with the observations, and Nuria Calvet for discussions about disk models, 
and for pointing us to other work underway on HD169142. 
We acknowledge NASA Origins of Solar Systems Program Grant NAG5-11777
for partial support.

\clearpage

\clearpage

\clearpage

\begin{deluxetable}{cc}
\tablewidth{0pt}
\tablecaption{HD169142 Observational Parameters}
\tablehead{
\colhead{Parameter} & \colhead{Value}}
\startdata
Observations      & 2005 April 19 (7 antennas)\\
Pointing center (J2000): &
  $\alpha=18^{h}24^{m}19\fs78$, $\delta=-29^{h}46^{m}49\fs37$ \\
Min/Max baseline   & 27 to 181 meters \\
Flux Calibrator    & Callisto \\
Primary Beam HPBW  & $55''$ \\
Synthesized Beam PBW   & $1\farcs6 \times 1\farcs0$, p.a. 20$^{\circ}$\\
K/Jy               & 15.4\\
Spectral Line setup:     & 512 channels/ 104 MHz \\
~~transition, frequency: & $^{12}$CO J=2--1, 230.538 GHz  \\
~~channel spacing:       & 0.27 km~s$^{-1}$ \\
r.m.s. (line images):    & 180 mJy \\
r.m.s. (continuum)  &  2.7 mJy \\
\enddata
\label{tab:obs}
\end{deluxetable}

\begin{deluxetable}{lccc}
\tablewidth{0pt}
\tablecaption{HD169142 Model Parameters} 
\tablehead{
\colhead{Parameter} &\colhead{Value}}
\startdata
Physical Structure & Irradiated accretion disk\tablenotemark{a} \\ 
Velocity Field     & Keplerian Rotation \\
Stellar Mass       & 2.0 M$_{\odot}$  \\
Turbulent Velocity & 0.12 km s$^{-1}$\\
Disk Outer Radius  & $235$ AU\\
Inclination Angle  & $13^{\circ}$ \\
Disk P.A.          & 5$^{\circ}$ \\
CO abundance       & $2.5\times 10^{-5}$ \\
\enddata
\tablenotetext{a}{\cite{dal05}}
\label{tab:model}
\end{deluxetable}


\begin{thebibliography}{}

\bibitem[Beckwith \& Sargent(1993)]{bec93} Beckwith, S.V.W. \& Sargent, A.I. 
  1993, \apj, 402, 280

\bibitem[Beckwith \& Sargent(1996)]{bec96} Beckwith, S.V.W. \& Sargent, A.I. 
  1996, Nature 383, 139

\bibitem[Beckwith et al.(1990)]{bec90} Beckwith, S.V.W., Sargent, A.I., 
  Chini, R., Gusten, R. 1990, \aj, 99, 924

\bibitem[Bouwman et al.(2001)]{bou01} Bouwman, J., Meeus, G., de Koter, A., 
  Hony, S., Dominik, C., Waters, L. B. F. M. 2001, \aap, 375, 950

\bibitem[Chiang et al.(2001)]{chi01} Chiang, E.I., Joung, M.K., 
  Creech-Eakman, M.J., Qi, C., Kessler, J.E., Blake, G.A., 
  van Dishoeck, E.F., 2001, \apj, 547 1077

\bibitem[Corder et al.(2005)]{cor05} Corder, S., Eisner, J., Sargent, A.
  2005, \apj, 622, L133

\bibitem[D'Alessio et al.(2005)]{dal05} D'Alessio, P., Merin, B., Calvet, N., 
  Hartmann, L., Montesinos, B. 2005, RevMex, 41, 61

\bibitem[Dent et al.(2005a)]{den05a} Dent, W. R. F., Greaves, J. S., 
  Coulson, I. M. 2005, \mnras, 359, 663

\bibitem[Dent et al.(2005b)]{den05b} Dent, W.R.F., Torrelles, J.M., 
  Osorio, M., Calvet, N., \& Anglada, G. 2005, MNRAS, submitted

\bibitem[Dominik et al.(2003)]{dom03} Dominik, C., Dullemond, C.P., 
  Waters, L.B.F.M., Walch, S. 2003, \aap, 398, 607

\bibitem[Dunkin et al.(1997)]{dun97} Dunkin, S.K., Barlow, M.J., 
  Ryan, Sean G. 1997, \mnras, 286, 604

\bibitem[Greaves et al.(2000)]{gre00} Greaves, J., Mannings, V., Holland, W. 
  2000, Icarus, 143, 155

\bibitem[Habart et al. (2005)]{hab05} Habart, E., Natta, A., Testi, L., 
  Carbillet, M. 2005, astro-ph/05031105

\bibitem[Hillenbrand et al.(1992)]{hil92} Hillenbrand, L.A., Strom, S.E., 
  Vrba, F.J., Keene, J. 1992, \apj, 397, 613

\bibitem[Ho et al.(2004)]{ho04} Ho, P.T.P., Moran, J.M., Lo, K.Y. 2004,
  \apj, 616, L1

\bibitem[Hogerheijde and van der Tak(2000)]{hog00} Hogerheijde, M. and 
  van der Tak, F. 2000, \aap, 362, 697

\bibitem[Kuhn et al.(2001)]{kuh01} Kuhn, J. R., Potter, D., Parise, B. 
  2001, \apj, 553, L189

\bibitem[Malfait et al.(1998)]{mal98} Malfait, K., Bogaert, E., Waelkens, C. 
  1998, \aap, 331, 211

\bibitem[Mannings \& Sargent(1997)]{man97} Mannings, V., Sargent, A. 1997, 
  \apj, 490, 792

\bibitem[Mannings \& Sargent(2000)]{man00} Mannings, V., Sargent, A. 2000, 
  \apj, 529, 391

\bibitem[Meeus et al.(2001)]{mee01} Meeus, G., Waters L. B. F. M., Bouwman, J., 
  van den Ancker, M. E., Waelkens, C., Malfait, K. 2001, \aap, 365, 476

\bibitem[Natta et al.(2004)]{nat04} Natta, A., Testi, L., Neri, R., 
  Shepherd, D., Wilner, D. 2004, \aaps, 416, 179

\bibitem[Pietu et al.(2003)]{pie03} Pietu, V., Dutrey, A., Kahane, C. 2003, 
  \aap, 398, 565

\bibitem[Pietu et al.(2005)]{pie05} Pietu, V., Guilloteau, S., Dutrey, A.,
  2005, \aap, in press (astro-ph/0504023)

\bibitem[Simon et al.(2000)]{sim00} Simon, M, Dutrey, A., Guilloteau, S.
  2000, \apj, 545, 1034

\bibitem[Sargent \& Beckwith(1987)] Sargent, A., Beckwith, S. 1987, 
  \apj, 323, 394

\bibitem[Sylvester et al.(1996)]{syl96} Sylvester, R.J., Skinner, C.J., 
  Barlow, M.J., Mannings, V. 1996, \mnras, 279, 915

\bibitem[Th\'{e} et al.(1994)]{the94} Th\'{e}, P.S., de Winter, D., 
  Perez, M.R. 1994, \aaps, 104, 315

\end{thebibliography}
\end{document}